\documentclass[%
aps,
prx,
reprint,
superscriptaddress,
nofootinbib,
amsmath,
amssymb
]{revtex4-2}
\usepackage{
    physics,
    graphicx,
    multirow,
    tabularx,
    mathtools, %for \begin{rcases} \end{rcases}
    placeins, %To include \FloatBarrier
    nicefrac, %For inline fraction
    hyperref, %To use put in url, enable links in bibiliography
    threeparttable, % For table footnotes
    siunitx,
    enumitem,
    makecell
}

\usepackage[capitalize]{cleveref}
\Crefname{section}{Sec.}{Secs.}

\usepackage[dvipsnames]{xcolor}
\hypersetup{
    colorlinks,
    linkcolor={blue!50!black},
    citecolor={blue!50!black},
    urlcolor={blue!80!black}
}% remove the box around citation link and bibilio link

\usepackage[caption=false]{subfig} %The caption part is not compatible with revtex
\graphicspath{{Figures/}}

\newcommand{\matinv}{{\boldsymbol{S}}^{-1}}
\newcommand{\mat}{{\boldsymbol{S}}}
\newcommand{\myvar}{\mathrm{Var}}

\newcommand{\expect}[1]{{\mathbb{E}[#1]}}
\renewcommand{\var}[1]{{\mathrm{Var}[#1]}}

\DeclareMathOperator*{\argmin}{arg\,min}

\usepackage{xparse}
\DeclareDocumentCommand\pbra{ s m }
{ % Bra
	\IfBooleanTF{#1}
	{\vphantom{#2}\left\langle\!\left\langle\smash{#2}\right\rvert\right.} % No resize
	{\left\langle\!\left\langle{#2}\right\rvert\right.} % Auto sizing
}
\DeclareDocumentCommand\pket{ s m }
{ % Ket
	\IfBooleanTF{#1}
	{\vphantom{#2}\left.\left\lvert\smash{#2}\right\rangle\!\right\rangle} % No resize
	{\left.\left\lvert{#2}\right\rangle\!\right\rangle} % Auto sizing
}
\DeclareDocumentCommand\pbraket{ s m g }
{ % Inner product
	\IfBooleanTF{#1}
	{ % No resize
		\IfNoValueTF{#3}
		{\vphantom{#2}\left\langle\!\left\langle\smash{#2}\middle\vert\smash{#2}\right\rangle\!\right\rangle}
		{\vphantom{#2#3}\left\langle\!\left\langle\smash{#2}\middle\vert\smash{#3}\right\rangle\!\right\rangle}
	}
	{ % Auto resize
		\IfNoValueTF{#3}
		{\left\langle\!\left\langle{#2}\middle\vert{#2}\right\rangle\!\right\rangle}
		{\left\langle\!\left\langle{#2}\middle\vert{#3}\right\rangle\!\right\rangle}
	}
}
\DeclareDocumentCommand\pketbra{ s m g }
{ % Dyad
	\IfBooleanTF{#1}
	{ % No resize
		\IfNoValueTF{#3}
		{\vphantom{#2}\left\lvert\smash{#2}\middle\rangle\!\middle\rangle\!\middle\langle\!\middle\langle\smash{#2}\right\rvert}
		{\vphantom{#2#3}\left\lvert\smash{#2}\middle\rangle\!\middle\rangle\!\middle\langle\!\middle\langle\smash{#3}\right\rvert}
	}
	{ % Auto resize
		\IfNoValueTF{#3}
		{\left\lvert{#2}\middle\rangle\!\middle\rangle\!\middle\langle\!\middle\langle{#2}\right\rvert}
		{\left\lvert{#2}\middle\rangle\!\middle\rangle\!\middle\langle\!\middle\langle{#3}\right\rvert}
	}
}

\begin{document}

%\preprint{APS/123-QED}

\title{High-Dimensional Subspace Expansion Using Classical Shadows}

\newcommand{\qmaddress}{\affiliation{Quantum Motion, 9 Sterling Way, London N7 9HJ, United Kingdom}}
\newcommand{\oxddress}{\affiliation{Department of Materials, University of Oxford, Parks Road, Oxford OX1 3PH, United Kingdom}}
\newcommand{\mathaddress}{\affiliation{Mathematical Institute, University of Oxford, Woodstock Road, Oxford OX2 6GG, United Kingdom}}

\author{Gregory Boyd}
\email{gregory.boyd@materials.ox.ac.uk}
\oxddress
\mathaddress

\author{B\'alint Koczor}
\mathaddress
\oxddress
\qmaddress

\author{Zhenyu Cai}
\email{cai.zhenyu.physics@gmail.com}
\oxddress
\qmaddress

\date{\today}% It is always \today, today,

\begin{abstract}
    We introduce a post-processing technique for classical shadow measurement data that enhances the precision of ground state estimation through high-dimensional subspace expansion; the dimensionality is only limited by the amount of classical post-processing resources rather than by quantum resources. Crucial steps of our approach are the efficient identification of ``useful'' observables from shadow data, followed by our regularised subspace expansion that is designed to be numerically stable even when using noisy data. We analytically investigate noise propagation within our method, and upper bound the statistical fluctuations due to the limited number of snapshots in classical shadows. In numerical simulations, our method can achieve a reduction in the energy estimation errors in many cases, sometimes by more than an order of magnitude. We also demonstrate that our performance improvements are robust against both coherent errors (bad initial state) and gate noise in the state-preparation circuits. Furthermore, performance is guaranteed to be at least as good---and in many cases better---than direct energy estimation without using additional quantum resources and the approach is thus a very natural alternative for estimating ground state energies directly from classical shadow data.
\end{abstract}

\maketitle

\section{Introduction}

Quantum computation often relies on the post-processing of measurement results from a quantum device to produce meaningful results, either for reducing the load on quantum hardware by offloading computation that can be performed classically~\cite{shorAlgorithmsQuantumComputation1994,peruzzoVariationalEigenvalueSolver2014,grinkoIterativeQuantumAmplitude2021}, for mitigating errors in noisy computations~\cite{caiQuantumErrorMitigation2023}, or for both~\cite{chenRobustShadowEstimation2021,huLogicalShadowTomography2022,jnaneQuantumErrorMitigated2024}. Classical shadow techniques~\cite{huangPredictingManyProperties2020} can output a rich set of classical data that is amenable towards various post-processing techniques for efficiently probing many properties of the quantum state, offering the ability to ``measure first, ask questions later''. Its versatility has been employed for a wide range of important tasks in quantum information \cite{chanAlgorithmicShadowSpectroscopy2023, boydTrainingVariationalQuantum2022, nakajiMeasurementOptimizationVariational2023, brydgesProbingRenyiEntanglement2019,struchalinExperimentalEstimationQuantum2021, abbasQuantumBackpropagationInformation2023}.

Ground state energy estimation is one of the most fundamental
tasks in quantum simulation~\cite{mcardleQuantumComputationalChemistry2020,daleyPracticalQuantumAdvantage2022}
and is also more widely applicable in optimisation problems~\cite{abbasQuantumOptimizationPotential2023}.
One typically aims to prepare a good approximation to the ground state through a state-preparation circuit, from which one can directly estimate its energy as the expected value of the problem Hamiltonian.
However, this often incurs an estimation error due to constraints in the size of the state preparation circuit and/or noise in the circuit. In this work, we present a post-processing method using existing classical shadow measurement data to produce more accurate ground state energy estimation by performing subspace expansions \cite{mccleanDecodingQuantumErrors2020,mccleanHybridQuantumclassicalHierarchy2017} of unprecedented dimension, mitigating the constraints due to both the circuit size and the noise in the shadow data.
Given our approach is completely performed in post processing, it does not require any additional quantum resources and circuit runs, and it is guaranteed to output a result that is at least as good as the original value, with the possibility of achieving orders of magnitude performance improvement, especially in the cases with a highly mismatched starting state and/or a large amount of shadow data. 

Our main contributions are a set of techniques for extracting relevant data from classical shadows and then performing numerically stable subspace expansions of high dimension (up to $10000$) using noisy data. We investigate the performance of such high dimensional subspace expansions using simulated classical shadows of up to 14 qubits, as well as simulations of the procedure with gate noise, demonstrating the effectiveness of our scheme as an error mitigation method. We also provide an analysis of noise propagation in the method as well as a derivation of the variances of the observables due to the finite number of snapshots from the classical shadow.

In \cref{sec:background} we provide background on the two main components of our work, subspace expansion and classical shadows, \cref{sec:methods} provides details on our shadow subspace expansion procedure and how we perform the numerical post-processing to ensure a well-conditioned eigenvalue problem. In \cref{sec:shot_noise_propagation}, we present our argument for the propagation of shot noise from the calculated matrix elements onto the rest of the computation, and in \cref{sec:results} we present the results of numerical simulations of our method, investigating the effect of shot noise from classical shadows as well as simulated gate noise.

\section{Background} \label{sec:background}
Here we first introduce the basic concepts that will form the core of our approach, such as subspace expansion using quantum computers and classical shadows for extracting a large number of expected values.
\subsection{Subspace Expansion}
In many applications of quantum computers  (e.g. variational eigensolvers) the aim is to directly prepare a quantum state that minimises a certain cost function. Let us consider the case where the cost function is the energy of the state with respect to a particular problem Hamiltonian $H_{\mathrm{tot}}$, which means that we are trying to prepare the corresponding ground state $\rho_{0}$. It is often the case that we cannot prepare the exact ground state $\rho_{0}$ due to limitations in qubit overhead, gate overhead or noise, and the best approximation of the ground state is some---potentially noisy---state that we denote as $\rho$. Quantum subspace expansion is a way to effectively improve on the observable expectation values obtained from $\rho$ by performing additional measurements and post-processing. 

In the case when $\rho$ is a pure state $\ketbra{\psi}$, we can improve upon $\ket{\psi}$ by constructing a subspace spanned by the set of states $\{G_i\ket{\psi} \ |\ 1\leq i \leq N_G\}$, which are states obtained by applying a set of \emph{expansion basis} operators $\{G_i\}$ to $\ket{\psi}$. We will then try to find the state within this subspace that minimises the energy. Note that $I$ is always an element of $\{G_i\}$ such that our starting state $\ket{\psi}$ is also a state within this subspace, thus the final state we obtain by definition has energy lower than or equal to that of $\ket{\psi}$. Any quantum state in this subspace can simply be written as 
\begin{align*}
    \ket{\psi_{\vec{w}}} = \frac{ \sum_{i = 1}^{N_{G}} w_i G_i \ket{\psi}}{\norm{ \sum_{i = 1}^{N_{G}} w_i G_i \ket{\psi}}} = \frac{\Gamma_{\vec{w}} \ket{\psi}}{\norm{\Gamma_{\vec{w}} \ket{\psi}}}
\end{align*}
where $ \Gamma_{\vec{w}} = \sum_{i = 1}^{N_{G}} w_i G_i$
is the \emph{expansion operator} which is parametrised by the weight vector $\vec{w}$. Different states $\ket{\psi_{\vec{w}}}$ in the subspace simply correspond to different weight vectors $\vec{w}$. For simplicity, in this article, we will assume that the set of operators $\{G_i\}$ are Hermitian, and the weight vectors $\vec{w}$ are real, which implies that $\Gamma_{\vec{w}}$ is also Hermitian. However, many of our arguments will be applicable to more general cases.

The weight vector (and the corresponding state) of minimal energy within this subspace is given as
\begin{align}\label{eqn:pure_state_expansion}
    \vec{w}^* = \argmin_{\vec{w}}  \bra{\psi_{\vec{w}}} H_{\mathrm{tot}} \ket{\psi_{\vec{w}}}.
\end{align}

Applying the same argument to the case where our starting state $\rho$ is a mixed state, the subspace-expanded state is then given as 
\begin{align*}
    \rho_{\vec{w}} = \frac{\Gamma_{\vec{w}} \rho \Gamma_{\vec{w}}}{\Tr(\Gamma_{\vec{w}}^2 \rho)},
\end{align*}
and the optimisation problem becomes
\begin{align}\label{eqn:mixed_state_expansion}
    \vec{w}^* = \argmin_{\vec{w}}  \Tr(H_{\mathrm{tot}} \rho_{\vec{w}}).
\end{align}

In order to obtain the weights $\vec{w}$ that extremises the energy $\Tr(H_{\mathrm{tot}}\rho_{\vec{w}})$, we need to solve the generalised eigenvalue problem:
\begin{align} \label{eq:generalised_eigenvalue_problem}
    \boldsymbol{H} \vec{w} = E_{\vec{w}}\boldsymbol{S}\vec{w}
\end{align}
where
\begin{align}\label{eqn:matrix_elem}
    \boldsymbol{H}_{ij} &= \Tr(G_i H_{\mathrm{tot}}G_j\rho)\\
    \boldsymbol{S}_{ij} &= \Tr(G_i G_j\rho).
\end{align}

If it is the ground state energy that we are interested in, then we obtain an improved ground state energy estimate $E_{\vec{w}^*}$ as the eigenvalue when we solve the equation above. If we are interested in the expectation value of some other observable $O$ with respect to the ground state, then we can measure the following matrix elements
\begin{align*}
\boldsymbol{O}_{ij} &= \Tr(G_i OG_j\rho)
\end{align*}
for different $i, j$ and then reconstruct the improved observable expectation value using:
\begin{align*}
    \Tr(O \rho_{\vec{w}^*}) &= \frac{1}{\Tr(\Gamma_{\vec{w}^*}^2 \rho)} \sum_{i,j = 1}^{N_G}w^*_i\Tr(G_{i} OG_j\rho )w^*_j \\
    &= \frac{\vec{w}^{ *\dagger} \boldsymbol{O} \vec{w}^{*}}{\Tr(\Gamma_{\vec{w}^*}^2 \rho)} .
\end{align*}

\subsection{Shadow Tomography}
We see that the expansion procedure requires us to measure the matrix elements $\boldsymbol{H}_{ij}$, $\boldsymbol{S}_{ij}$ and $\boldsymbol{O}_{ij}$, which is $\order{N_{G}^2}$ observables that we need to measure. It is also often the case that the Hamiltonian cannot be directly measured and needs to be broken down to $N_{H}$ sub-terms, which means that the total number of observables that we need to measure is $\order{N_{G}^2N_{H}}$. Therefore, it can be costly to further increase the dimension of the expansion space $N_G$ to achieve better subspace expansion results. One way to efficiently measure multiple observables is to use shadow tomography~\cite{aaronsonShadowTomographyQuantum2018,huangPredictingManyProperties2020}. To construct a classical shadow of a quantum state, we repeat the following process: we sample a random unitary $U_i$ from a distribution of unitaries $\mathbb U$, apply it to the quantum state and then measure the resulting state in the computational basis. These steps combined can be described as the \textit{process channel} $\mathcal M$. For certain choices of $\mathbb U$, we are able to efficiently invert the channel from the output bit string $\ket{\boldsymbol{b}}$ classically to obtain a classical snapshot $\rho_l = \tilde{\mathcal{M}} (U_i \ketbra{\boldsymbol{b}} U_i^\dagger)$, where $\tilde{\mathcal M}$ is the classical inversion of the process channel, and then use these snapshots to estimate observables in post-processing. 

In this work, we primarily consider the case of Pauli shadows, where $\mathbb U$ consists of all the tensor products of single-qubit Clifford operators. Using this Pauli shadow scheme, the sample complexity of obtaining estimators of $M$ Pauli operators of weight $l$ to error $\epsilon$ is $\mathcal O (3^l \log(M )/\epsilon^2)$~\cite{huangPredictingManyProperties2020}, meaning that provided both the operators in the set $\{G_i\}$ and the terms in the Hamiltonian are local, we can efficiently determine many expectation values to perform high-dimensional subspace expansion. Do note that the $\log(M)$ factor above exists to make sure \emph{all} $M$ estimators reach the stated precision $\epsilon$. In our case, our end goal is not estimating these $M$ observables individually; instead, they are used for further post-processing to predict one single observable of interest, thus this $\log(M)$ factor is not applicable to our case.

\section{Methods} \label{sec:methods}
We now set out the methodology for Shadow Subspace Expansion (SSE) and some practical considerations
in performing large subspace expansions in post-processing using classical shadow data.
While the present approach is immediately compatible with more advanced shadow protocols like shallow shadows~\cite{bertoniShallowShadowsExpectation2022} or fermionic shadows~\cite{wanMatchgateShadowsFermionic2023}, for simplicity, we will demonstrate and explain our results on 
the example of local Pauli shadows detailed above
-- which is very efficient
for simultaneously estimating low-weight Pauli observables.

For this reason we construct the expansion operators $\Gamma_{\vec{w}^*}$ as linear combinations of
local Pauli strings. The central assumption of subspace expansion is that the expansion basis will map the starting state $\rho$ into another basis state in the subspace that has significant components orthogonal to the starting state, increasing the effective dimension of our subspace. For our special case in which the expansion basis is the set of local Pauli strings, this can be especially effective in many practical scenarios as discussed below and demonstrated later in our numerical experiments. 

One such application scenario is the translation invariant, gapped spin problems,
whereby elementary excitation operators are supported on local operators~\cite{haegemanElementaryExcitationsGapped2013}
(with an exponentially decreasing error as the locality increases).
As such, given a jump operator that maps the ground state to the first excited state as $G\ket{0} = \ket{1}$
and vice versa, any low energy eigenstate of the form $\ket{\psi} = \sqrt{1-\epsilon} \ket{0} +  \sqrt{\epsilon} \ket{1}$
can be mapped to the ground state as  $\ket{0} =  \Gamma_{\vec{w}^*} \ket{\psi }$ with $\Gamma_{\vec{w}^*} = \sqrt{1-\epsilon} \openone -  \sqrt{\epsilon} G $ and some suitable normalisation.
Indeed, for more complex initial states one potentially needs significantly more expansion terms, however,
a low-energy initialisation guarantees high overlap with the low-lying eigenstates.

Furthermore, in fermionic systems the ground state $\ket{0}$ is typically obtained as
an expansion of the HF state $\ket{\psi_{HF}}$ into low-weight fermionic annihilation-creation operators.
Ref.~\cite{chanAlgorithmicShadowSpectroscopy2023} details that after encoding the fermionic problem as a qubit Hamiltonian via the 
Jordan-Wigner transform, this expansion has a significant overlap with local Pauli operators
despite the fermionic operators being mapped to non-local Pauli operators.

\subsection{Shadow Subspace Expansion Procedure} \label{sec:procedure}
We are considering the setting in which we want to estimate some property $O_{\mathrm{tot}}$ of the ground state $\rho_0$ of some Hamiltonian $H_{\mathrm{tot}}$, with their Pauli decompositions being $O_{\mathrm{tot}} = \sum_{k = 1}^{N_{O}} O_k$ and $H_{\mathrm{tot}} = \sum_{k = 1}^{N_H} H_k$, respectively. This can be done by preparing $\rho_0$ and obtaining a set of its classical shadows of size $N_s$ for the estimations of various properties through post-processing. However, due to noise in the circuit and/or imperfect ansatz circuit, the state we prepare and the shadow data we obtain are from some other state $\rho$ instead. We will now outline the way to obtain an improved estimate of $\Tr(O_{\mathrm{tot}}\rho_0)$ using the shadow data of $\rho$.

\begin{enumerate}[itemsep=0pt,topsep=4pt,leftmargin=3ex]

    \item \textbf{Initial expansion basis:} Select an initial expansion basis $\{G_i\}$ of size $N_{G}$. One way is to use the set of all Pauli strings up to a certain weight, where the weight is determined by the amount of shadow data (the more shadow data, the higher the weight we can reach). 
    \item \textbf{Filtering expansion operators:} For each expansion basis operator $G_i$, we will perform a local subspace expansion of dimension $2$ using the expansion basis $\{I, G_i\}$ and the corresponding improvement in energy is denoted as $\Delta E_i$. This involves determining the expected values of the observables $\{G_i\}$, $\{H_k\}$ and $\{G_i H_k G_i\}$ (which are just the $H_k$ with an added phase as the $G_i$ can be (anti)commuted through the $H_k$). These observables have weights up to $M_G, M_H$, respectively,
    and overall the total number of expectation values to be determined from the measurement data is $N_{G}+ N_H$.
    
    To filter out expansion basis operators that are unable to
    decrease the energy of the initial state, we will rank all operators in $\{G_i\}$ according to their improvement in energy $\Delta E_i$ and keep only the top $K$ operators, which gives us the \emph{$K$-significant} expansion basis. This process has a linear classical computational complexity of $O(N_G N_H)$; thus, $N_G$ can be chosen as potentially very large, e.g., $N_G \gtrsim 10^6$.

    \item \textbf{Equation for full expansion:} We will perform subspace expansion using the $K$-significant expansion basis, which involves estimating the set of weight-$2M_{G}$ observables $\{G_iG_j\}$ to form the matrix $\boldsymbol{S}$, and weight-$2M_{G}+M_{H}$ observables $\{G_iH_kG_j\}$ to form the matrix $\boldsymbol{H}$, the matrix entries can be determined based on the observation that any product of local Pauli strings evaluates to a single Pauli string of higher locality~\cite{boydTrainingVariationalQuantum2022}. They require the estimations of $K^2$ and $K^2N_{H}$ observables, respectively.

    \item \textbf{Solving the eigenvalue problem:} To obtain the eigenspectrum in the subspace and the optimal weights $\vec{w}$. We detail our method in \cref{sec:truncation_rule}
    for solving the noisy generalised eigenvalue problem in \cref{eq:generalised_eigenvalue_problem}.
    This step has a classical computational complexity of $O(K^p)$ with a polynomial power $2 \leq p \leq 3$
    and is thus the time bottleneck. This reflects the importance of the screening step 
    for choosing the best $K$ observables.

\end{enumerate}

 Step 4 will output the energy of the state found via subspace expansion. The following steps are needed for estimating other observables on the output state:

\begin{enumerate}
[itemsep=0pt,topsep=4pt,leftmargin=3ex]
  \setcounter{enumi}{4}
    \item Use the shadows to estimate the set of weight-$2M_{G}+M_{O}$ observable $\{G_iO_kG_j\}$ of a size $K^2N_{O}$. 
    \item Use the optimal weight $\vec{w}^*$ to reconstruct the optimal observable
    \begin{align*}
        \Tr(O_{\mathrm{tot}}\rho_{\vec{w}^*}) &= \frac{\Tr(O_{\mathrm{tot}}\Gamma_{\vec{w}^*}\rho\Gamma_{\vec{w}^*})}{\Tr(\Gamma_{\vec{w}^*}\rho\Gamma_{\vec{w}^*})}\\
        % &= \frac{\Tr(\Gamma_{\vec{w}^*}O_{\mathrm{tot}}\Gamma_{\vec{w}^*}\rho)}{\Tr(\Gamma_{\vec{w}^*}^2\rho)}\\
        &= \frac{\sum_{i,j} \beta_k w_i^*w_j^* \Tr(G_iO_kG_j\rho)}{\Tr(\Gamma_{\vec{w}^*}^2\rho)}
    \end{align*}
\end{enumerate}

We note that there are alternative methods for choosing the $K$-significant observables and we have described one such method based on Shadow Spectroscopy~\cite{chanAlgorithmicShadowSpectroscopy2023} in \cref{sec:shadow_spectroscopy_observables}.

\begin{figure*}[htbp]
	\centering
	\subfloat[ \label{fig:errorplotLow_spinchain}]{\includegraphics[width=0.45\linewidth]{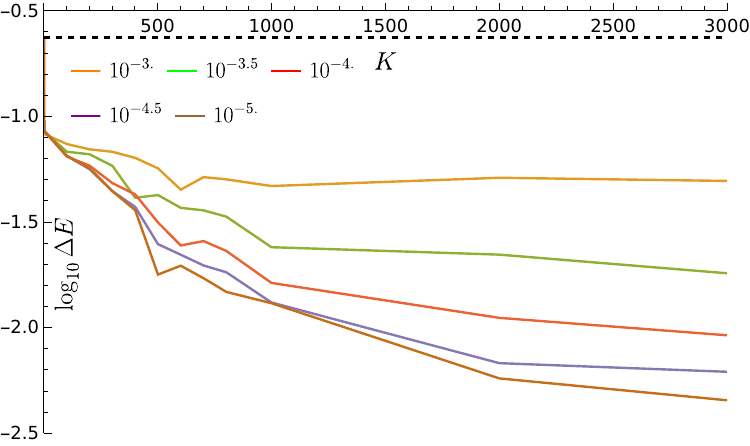}} \quad \quad \subfloat[\label{fig:errorplot_spinchain}]{\includegraphics[width=0.45\linewidth]{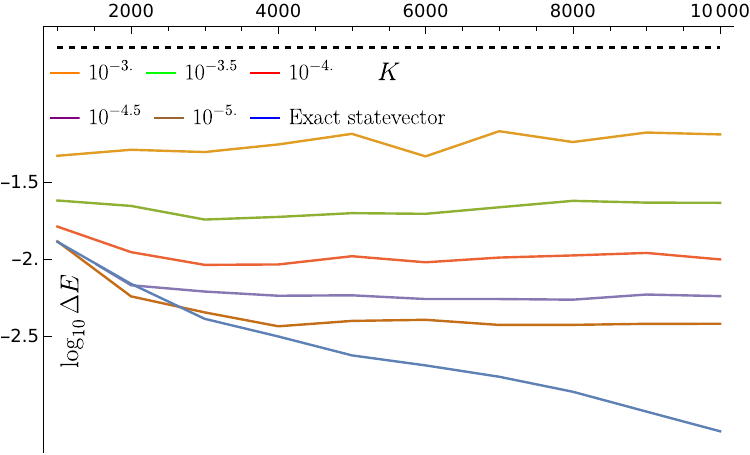}}\\
	\subfloat[\label{fig:SVs_vs_noise}]{\includegraphics[width=0.45\linewidth]{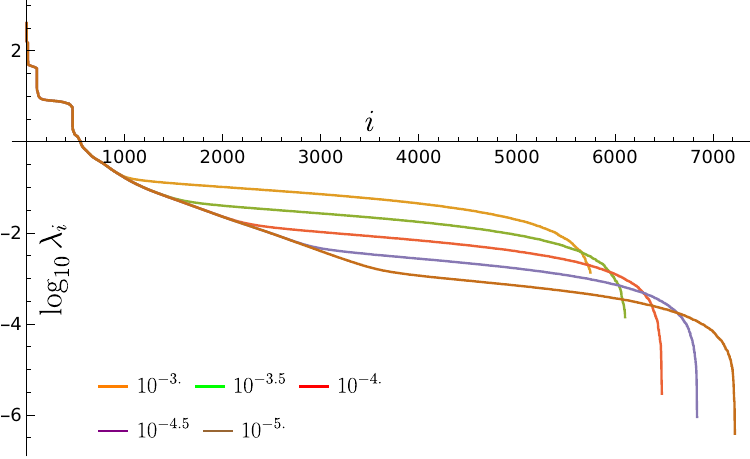}}
	\quad\quad
	\subfloat[\label{fig:SVs_vs_K}]{\includegraphics[width=0.45\linewidth]{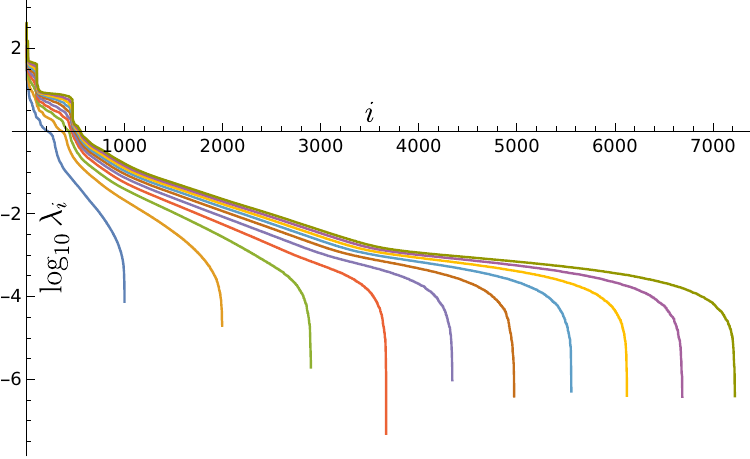}}
	\caption{(a) Error of SSE for the spin chain problem with subspace dimension $K$, with shot noise level $\epsilon$ per matrix element (denoted by the legend) with the energy of the input state denoted by the black dotted line. (b) Errors as $K$ increases demonstrating that increasing $K$ beyond a threshold set by the noise level will not decrease the estimation error.
		(c) Impact of matrix element error on the eigenvalues of the matrix $\boldsymbol{S}$, for $K=10000$ on the spin chain problem (with unphysical negative eigenvalues discarded), indicating that the source of the decrease in accuracy in energy is the removal of information from the low magnitude eigenvalues. The truncation dimension required is typically found to be well above the noise-induced levelling off on this plot. (d) Impact of increasing $K$ on the eigenvalues of the matrix $\boldsymbol{S}$ for $\epsilon=10^{-5}$ indicating that the significant eigenvalues are increased as more information in included in the subspace expansion by increasing $K$, however this does not substantially raise lower magnitude eigenvalues above the noise floor.
	} 
\end{figure*}

\subsection{Shot noise propagation} \label{sec:shot_noise_propagation}
As explained in the previous section, we proceed by estimating matrix entries of $\boldsymbol{S}$
and $\boldsymbol{H}$ from classical shadows data. Due to the finite sample size, the matrix entries come with a certain amount of shot noise. 
Building on the main theorem of ref.~\cite{vanstraatenMeasurementCostMetricAware2021},
we bound how this shot noise propagates into the solution of the generalised subspace expansion,
which helps us pinpoint the hyperparameters that require particular attention in the post-processing step.

We can recast the generalised eigenvalue equation in \cref{eq:generalised_eigenvalue_problem} into the usual eigenvalue equation $\boldsymbol{M} \vec{w} = E_{\vec{w}} \vec{w}$ where $\boldsymbol{M} = \matinv\boldsymbol{H}$.
We quantify the shot noise in the matrix $\boldsymbol{M}$ as
the total variance $\epsilon_{M}^2 = \sum_{kl} \myvar\big\{ [\boldsymbol{M}]_{kl} \big\}$ which is equivalent
to the expected Hilbert-Schmidt distance of the matrix due to shot noise. We obtain the following bound in \cref{app:proof_propagation} as
\begin{equation}\label{eq:error_prop}
	\epsilon_{M}^2 \leq  N_s^{-1} \, 3^{2w'}  K  \Big[   \lVert H_{\mathrm{tot}} \rVert_\infty  \lVert \matinv \rVert_2^4  +3^{w } \lVert \matinv \rVert_2^2  \Big ].
\end{equation}

While the above bound is based on worst-case arguments and is thus expected to be very pessimistic, it nevertheless, informs us of the following. 
First, the total variance can be reduced inversely proportionally by increasing the number of shots $N_s$ (standard shot noise scaling). 
Second, as we restrict ourselves to simple Pauli shadows \cite{huangPredictingManyProperties2020} we have an exponential dependence on the largest weight $w'$ among the expansion operators $G_k$ and on
 the largest weight among the Pauli terms in the total Hamiltonian $H_{\mathrm{tot}}$ --  of course, more advanced shadow techniques~\cite{bertoniShallowShadowsExpectation2022, caiBiasedEstimatorChannels2024,kingTriplyEfficientShadow2024}
 can help mitigate this dependence.
Third, the effect of shot noise depends polynomially on the Hilbert-Schmidt norm  of the 
inverse matrix $\lVert \matinv \rVert_2$ and thus truncation to the most significant singular values, as we detail below in \cref{sec:truncation_rule}, is essential to ensure this matrix norm is well controlled.
Finally, the variance is also influenced by the absolute largest eigenvalue $\lVert H_{\mathrm{tot}} \rVert_\infty $ of the problem Hamiltonian $H_{\mathrm{tot}}$
which is indeed expected to grow polynomially with the system size, i.e., the number of qubits, in typical scenarios.

As we detail in \cref{app:shot_noise2}, our conclusion is consistent with  the analysis of Ref.~\cite{yoshiokaGeneralizedQuantumSubspace2022}
In particular, Ref.~\cite{yoshiokaGeneralizedQuantumSubspace2022} assumed that in a single experiment
the deviation due to shot noise from the ideal matrices $\boldsymbol{H}$ and $\boldsymbol{S}$ are expressed
as $\delta \boldsymbol{H}$ and $\delta \boldsymbol{S}$, respectively. Then, the deviation in the solution of the generalised eigenvalue equation
can be bounded as  $\delta E_{\vec{w}} \leq  \norm{\boldsymbol{S}^{-1}}_{\infty}  \left(\norm{\delta \boldsymbol{H}}_{\infty} +\lVert H_{\mathrm{tot}} \rVert_\infty \norm{\delta \boldsymbol{S}}_{\infty}\right)$.
Indeed, the same spectral properties appear in this bound, however, \cref{eq:error_prop} expresses a statistical variance due to propagation
as opposed to the noise propagation in a single sample -- hence \cref{eq:error_prop} explicitly depends on further parameters, such as the number of shots.

\subsection{Regularisation of the Generalised Eigenvalue Problem} \label{sec:truncation_rule}
Classical shadows allow us to efficiently estimate a very large number of expansion operators, however,
many states obtained through low-weight expansion operators have high overlaps. As a result, our generalised eigenvalue equation potentially involves highly singular matrices. As detailed the previous section, shot noise propagation is potentially magnified by these small singular values of the overlap matrix. Therefore, our approach requires careful regularisation.

In order to numerically solve the generalised eigenvalue problem in \cref{eq:generalised_eigenvalue_problem}, we must have a well-conditioned $\boldsymbol{S}$. In our case, the physical conditions also require $\mathbf{S}$ to be positive semi-definite. Therefore, we can regularise the problem by transforming the matrices $\boldsymbol{H},\boldsymbol{S}$ to a subspace in the state expansion space where $\boldsymbol{S}$ has sufficiently large positive eigenvalues. This is analogous to the process of projecting onto the nearest (in terms of Frobenius norm) positive semi-definite matrix~\cite{highamComputingNearestSymmetric1988}, except we also dynamically choose a threshold to remove some eigenvalues above $0$ to ensure the eigenvalue problem is well-conditioned. More exactly, the regularisation is performed by restricting the problem to the subspace spanned by the first $\tilde{K}$ eigenvectors of $\boldsymbol{S}$ ranked by eigenvalues, which is in a sense the effective dimension of the matrix containing information that is distinguishable above the noise. These eigenvectors produce a $K \times \tilde{K}$ transformation matrix $\boldsymbol{Q}$, which can be used to transform the original problem to a dimension-$\tilde K$ generalised eigenvalue problem with the following matrices:
\begin{align} \label{eq:matrix_transform}
	\tilde{\boldsymbol{S}} = \boldsymbol{Q}^\dagger \boldsymbol{S} \boldsymbol{Q},
 \quad \quad \text{and} \quad \quad
	\tilde{\boldsymbol{H}} = \boldsymbol{Q}^\dagger \boldsymbol{H} \boldsymbol{Q},
\end{align}
which outputs the energy $E_0^{\tilde{K}}$.

The optimal value of $\tilde{K}$ to be used can be determined from a series of such transformations by calculating the ground state energy $E_0^l$ for each and stopping when the energies become unstable. This is determined by tracking the variance of the energy estimates within a moving window the and terminating when the moving variance reaches a minimum, indicating a series of gradually varying energies as the regularisation reaches the optimal value before the eigenvalue problem becomes unstable, resulting in an increase in the moving variance.

\section{Results} \label{sec:results}
We will first examine the use of Shadow Subspace Expansion (SSE) for solving the ground state energy for a 14-qubit spin chain model. The spin chain Hamiltonian takes the form
\begin{equation}\label{eq:spin-ring}
    \mathcal{H}=J\sum_{i=1}^N \Vec{\sigma}_i \cdot \Vec{\sigma}_{i+1} + c_i Z_i.
\end{equation}
with $J=0.1$ and $c_i \in [-1,1]$ as uniformly random on-site energies. This Hamiltonian has been considered in studies of self-thermalization and many-body
localization and has been identified as a promising target for quantum advantage \cite{childsFirstQuantumSimulation2018}.\\
For the sake of demonstration, we initialise using the Variational Quantum Eigensolver (VQE) \cite{peruzzoVariationalEigenvalueSolver2014} for $300$ steps to an initial state with energy error $\Delta E = 0.235$ from the ground state, and then perform SSE. 

\subsection{Performance with increasing subspace size}

In this section, we will investigate the performance of SSE as the size of the subspace $K$ increases. For simplicity, we will first assume the shot noise that is added to all the observables that form the entries of $\boldsymbol{H}$ and $\boldsymbol{S}$ is drawn from a complex Gaussian distribution of mean $0$ and standard deviation $\epsilon$, and then examine the effects of accurately modelled shot noise from classical shadow data in \cref{sec:simulations_using_shadow_variances,sec:gate_noise}.

Previously in the standard application of subspace expansion, the cost of measuring the expectation values grows quadratically with the subspace size, thus we are usually restricted to very small subspace size of $K \sim 10$. There is no such restriction for SSE, in \cref{fig:errorplotLow_spinchain}, we apply SSE up to a subspace size of $K=3000$ with different levels of shot noise. We see that increasing $K$ beyond the $K \sim 10$ regime leads to a strong reduction in the errors in the estimated energy. We can achieve almost an order of magnitude reduction with a modest shot noise of $\epsilon \sim 10^{-3}$, and almost two orders of magnitude reduction when we have a low shot noise of $\epsilon \sim 10^{-5}$.

In \cref{fig:errorplot_spinchain}, we look at even larger subspace sizes up to $K=10^4$, which nears the maximum $K$ available when using up to $3$-local Paulis as our expansion basis operators ($10690$ for $14$ qubits). In the noiseless case, the error in the estimated ground state energy decreases monotonically with the increase of $K$. With a finite level of shot noise per matrix element, the error in the ground state energy found by SSE hits an \emph{error floor} as we increase the size of the subspace $K$. This error floor will decrease as we decrease the amount of shot noise, with a rough fit of $\Delta E_{\mathrm{floor}} \propto \epsilon^{0.6}$ found in our numerics.

The phenomenon of the $K$-independent error floor shown in \cref{fig:errorplot_spinchain} can be explained by looking at the eigenvalues of $\boldsymbol{S}$. In \cref{fig:SVs_vs_noise}, we show the eigenvalue spectrum of $\boldsymbol{S}$ for $K=10^4$ at varying levels of shot noise, demonstrating that the magnitude of some eigenvalues will be so small such that they are buried under the shot noise as the noise increases. The information about the subspace carried by these eigenvalues and their associated eigenvectors needs to be excluded in the subspace expansion to keep the generalised eigenvalue problem well-conditioned. Hence, for a given shot noise level, there is a point where increasing $K$ will only add low-magnitude eigenvalues buried under the noise and will not assist in improving the ground state energy, reaching the error floor mentioned. 

We also examine the effect of increasing $K$ on the eigenvalues of $\boldsymbol{S}$. In \cref{fig:SVs_vs_K}, we plot these eigenvalues for different $K$ at the noise level of $\epsilon=10^{-5}$. For low $K$, the fraction of eigenvalues that are above the level of the noise is greater, whereas for higher $K$, we are including an increasing number of operators that add little information to the subspace expansion, and so have a growing number of eigenvalues that only contain information about the noise. One thing to note is that as $K$ increases, the eigenvalues of the matrix grow roughly linearly (see \cref{sec:largest_SV_with_K}) but the eigenvalues fall off exponentially with the index $i$ (at least in the intermediate region between the large eigenvalues and those which consist only of noise), the linear increase in signal-to-noise ratio as $K$ increases is not able to raise many more eigenvalues above the noise threshold.

\begin{figure}
	\includegraphics*[width=\linewidth]{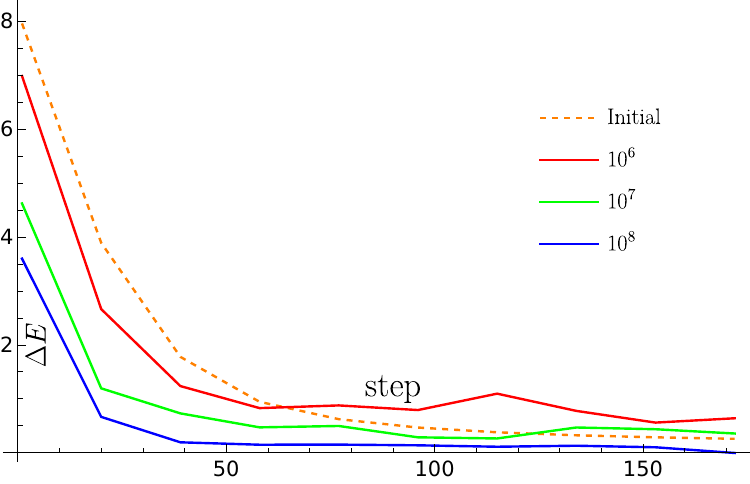}
	\caption{Energies of states over the course of a VQE optimisation of the spin chain model. The orange dashed line corresponds to the energy of the state at that point in the optimisation, whereas the red, green and blue curves correspond to the energies achieved using SSE (with $K=5000$) using accurate shadow variances for a shadow with the number of snapshots indicated in the legend.}
	\label{fig:direct_vs_subspace}
\end{figure}

\subsection{Simulations Using Shadow Variances} \label{sec:simulations_using_shadow_variances}

Until this point, we have been considering shot noise producing uniform Gaussian errors on the matrix elements of $\boldsymbol{S}, \boldsymbol{H}$, analogous to shot noise from measuring each of the elements independently. However, in the realistic setting where the expectation values are estimated from a fixed number of classical shadow snapshots, we do not expect the level of noise on each matrix element to be the same. In particular, when using local Clifford shadows, the variance of our estimators will have an extra factor of $3^w$, where $w$ is the weight of the observable (see \cref{sec:variance_local_clifford_shadow}). We therefore perform additional numerics where we accurately simulate the shot noise produced by the finite size of the classical shadow to examine its effect on the subspace expansion.
In \cref{fig:direct_vs_subspace}, we perform SSE on a series of states obtained from different steps along the VQE optimisation of the spin chain model, with the numbers of snapshots between $10^6$ and $10^8$, and compare the results to the energy of the initial state. We find substantial improvements for all numbers of snapshots for high energy initial states, but for states closer to the ground state, below a certain number of snapshots, the effect of shot noise on SSE means that we can get energy estimates with higher variance than direct estimation. This effect can be mitigated by increasing the number of snapshots. In our experiment, with $10^8$ snapshots, we will always obtain a better result compared to direct estimation.

\begin{figure*}[htbp]
	\centering
	\subfloat[ \label{fig:spinchain_gate_noise}]{\includegraphics[width=0.45\linewidth]{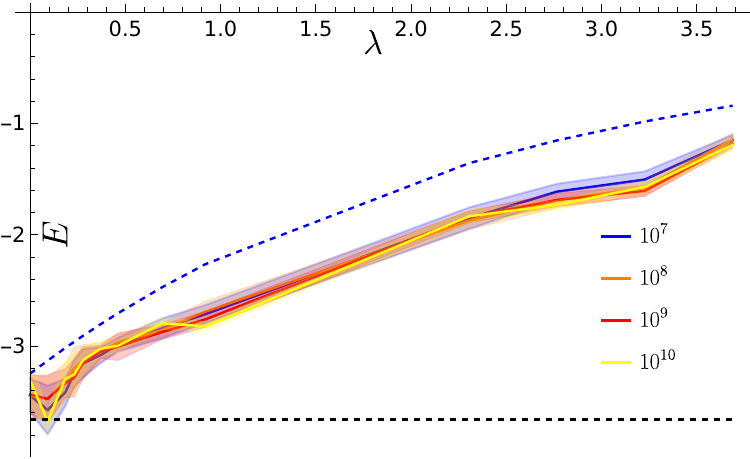}} \quad \quad \subfloat[ \label{fig:spinchain_gate_noise_delta}]{\includegraphics[width=0.45\linewidth]{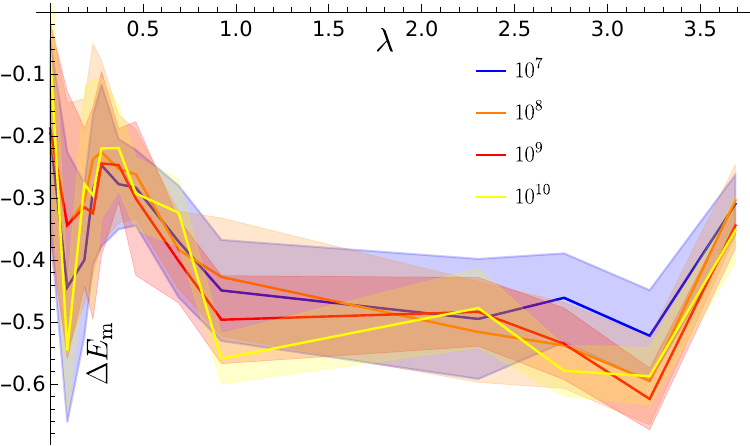}} \\
	\subfloat[\label{fig:spinchain_gate_noise_no_clamp}]{\includegraphics[width=0.45\linewidth]{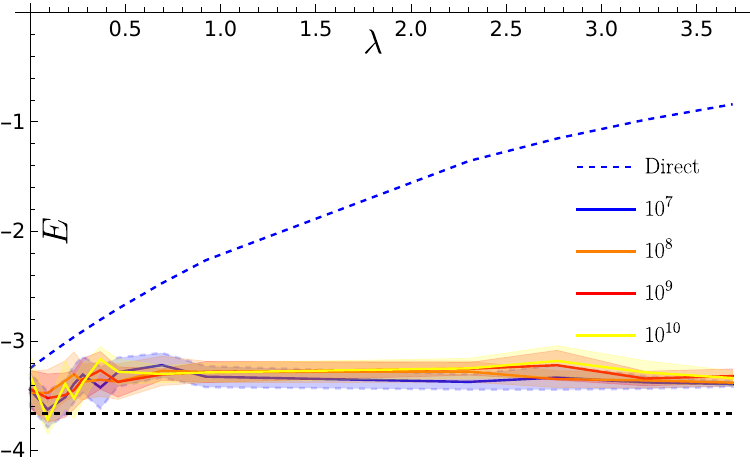}} \quad \quad \subfloat[\label{fig:spinchain_eff_dim_no_clamp}]{\includegraphics[width=0.45\linewidth]{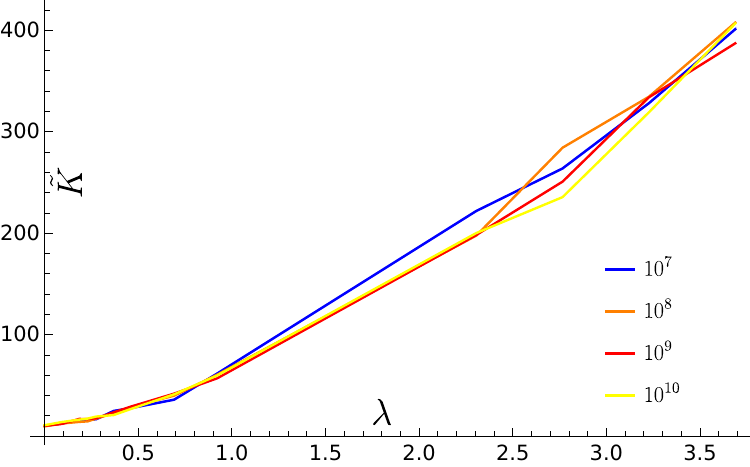}}
	\caption{(a) Energies obtained via simulations with gate noise, comparing the direct measurement of energy with the use of SSE with varying level of gate noise, denoted by the circuit fault rate $\lambda$ and shot noise from snapshot counts as given in the legend. SSE lines correspond to an average over $20$ separately sampled instances of shot noise with outliers removed, the shaded regions are the standard deviations of these distributions. The effective dimension of the subspace $\tilde K$ is allowed to be $15$ at maximum.
		(b) Energy improvements of the SSE mitigated observables over the direct estimation on the noisy state in \cref{fig:spinchain_gate_noise} showing an energy improvement of $\sim 0.5$ over the whole noise range, approximately independent of shot count.
		(c) The same calculation, as \cref{fig:spinchain_gate_noise} but with $\tilde K$ allowed to vary freely. The average values of $\tilde K$ found by the regulariser for the numerics in this plot are shown in \cref{fig:spinchain_eff_dim_no_clamp}.
		(d) Average value of $\tilde K$ found for the circuit used to generate \cref{fig:spinchain_gate_noise_no_clamp} with fault rate $\lambda$. The value of $\tilde K$ increases past the dimension of the Hilbert space in this small example, but remains well below the $4^n$ possible Pauli terms that can be used in the expansion for gate noise mitigation.}
\end{figure*}

It is worth noting that the original state (corresponding to the identity expansion operator) is in the full $K$-dimensional subspace. However, when we try to regularise the subspace via transformation and truncation as outlined in \cref{sec:truncation_rule}, the resultant $\tilde{K}$-dimensional subspace may not fully contain the original state anymore. This is the reason why we can reach an energy above direct estimation. This can be potentially solved by creating a new regularisation procedure explicitly keeping the original state in the output subspace, e.g. by regularising only the subspace outside the original state. In the present case, since the direct estimated energy, being an entry in the $\boldsymbol{H}$, is always a known quantity, we will simply output the direct estimated energy if the SSE energy is higher. In this way, we can ensure our method will always provide a better or at least an equally good estimate compared to direct estimation. 

Whilst we realise that finding a lower energy than direct estimation using SSE requires either a high energy starting state or a large number of snapshots, we note that there are multiple techniques that could alleviate the shot burden. For example, derandomisation \cite{huangEfficientEstimationPauli2021} can be used after the identification of good observables for the subspace expansion. We also note the existence of alternative classical shadow techniques that can achieve greater performance on some restricted classes of observables \cite{ippolitiClassicalShadowsBased2023} that could reduce the overhead based on problem-specific knowledge. Recent work also allows for the reduction of mean-squared error of the estimators produced from classical shadows via a bias-variance trade-off \cite{caiBiasedEstimatorChannels2024}, which would be effective at reducing the error on the high-weight observables.

\subsection{Effects of Gate Noise} \label{sec:gate_noise}

To investigate the effect of gate noise on our algorithm, we perform density matrix simulations on a 7-qubit spin chain, using a parametrised quantum circuit with $161$ single-qubit gates and $60$ two-qubit gates, adding single or two-qubit depolarising noise after each gate, with the noise strength for two-qubit gates being $5$ times of that for single-qubit gates.
We do this for a range of gate noise strengths leading to differing circuit fault rates $\lambda$, which are the expected number of faults per circuit execution, and simulate shot noise as in \cref{sec:simulations_using_shadow_variances} for a range of shot noise strengths denoted by the number of snapshots taken. We take $K=500$ and in \cref{fig:spinchain_gate_noise} show the performance of SSE on noisy circuits, producing an energy improvement over the direct noisy estimate of approximately $0.5$ over the noise range, with little dependence on the shot count as shown in \cref{fig:spinchain_gate_noise_delta}, demonstrating the robustness of the method as a form of error mitigation.

Noisy simulations of quantum circuits are resource intensive and thus we have only simulated up to $7$-qubits, which corresponds to a Hilbert space dimension of $d = 2^7$ with $256$ real parameters. The Hermitian and trace-preserving operator space that density operators live in has the dimension of $d^2-1 \sim 15000$. The small system size runs the risk of having an expansion subspace with a dimension larger than the degree of freedom needed to simulate the full quantum system. Hence, we have avoided this by clamping the maximum possible value of $\tilde K$ to be $15$ in \cref{fig:spinchain_gate_noise}. In \cref{fig:spinchain_gate_noise_no_clamp} we show the same simulations without any restrictions on $\tilde K$, and show the corresponding values of $\tilde K$ found in \cref{fig:spinchain_eff_dim_no_clamp}, where they increase with the noise level. We see that indeed increasing the dimension of the subspace is an excellent way to mitigate gate noise in the circuit. In fact as shown in \cref{fig:spinchain_gate_noise_no_clamp}, we are able to achieve an estimation accuracy that is largely independent of the circuit noise level if we put no restriction on the dimension of the expansion.

\subsection{Quantum Chemistry and Excited States} \label{sec:LiH_results}

\begin{figure}
    \includegraphics*[width=\linewidth]{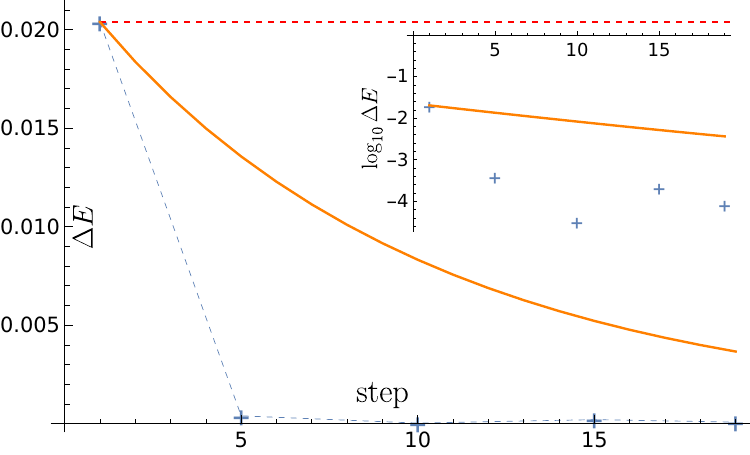}
    \caption{Energy estimation errors of direct estimation (orange curve) and SSE (blue markers) as a function of the number of VQE steps (orange, using UCCSD ansatz with a single layer) used to generate the initial expansion state, starting from the Hartree-Fock state for LiH. The Hartree-Fock energy is shown by the red dotted line, and SSE is unable to improve on this due to the long Pauli strings require but even after a small number of VQE steps, SSE is able to rapidly make improvements on the ground state energy. The inset shows the log of the energy error for SSE. Shot noise level $\epsilon=10^{-6}$.}
    \label{fig:LiH_initStates}
\end{figure}

Besides spin chain Hamiltonians, we also examine the performance of SSE on a quantum chemistry Hamiltonian. We use the 12-qubit LiH Hamiltonian in the STO-3G basis \cite{mccleanOpenFermionElectronicStructure2020}, and run VQE using a UCCSD ansatz \cite{leeGeneralizedUnitaryCoupled2019}. We then perform SSE on the output VQE states along the optimisation trajectory. In \cref{fig:LiH_initStates}, we show the VQE energies achieved by the orange line and the SSE energies achieved from the state at that step number using blue markers, with a shot noise of $\epsilon=10^{-6}$. SSE is unable to improve on the Hartree-Fock state due to its restriction to local Pauli strings, but even after a small number of steps, two orders of magnitude improvement in the energy estimation error $\Delta E$ can be achieved over the directly estimated VQE energy. In this case, the Hamiltonian of LiH used had a ground state with the correct electron number, and so no accommodation needed to be made for the electron number symmetry of the Hamiltonian. However, if this were not the case and the Hamiltonian had a `non-physical' ground state with a different number of electrons, the techniques in \cref{sec:symmetry} would be required to ensure the solution was not corrupted with lower energy states of the wrong electron number.

\begin{figure}
    \includegraphics*[width=\linewidth]{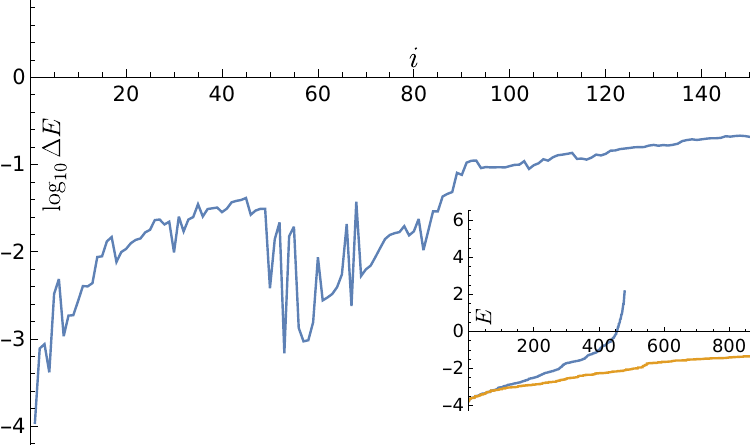}
    \caption{The energy difference between the spectrum produced by SSE and the true spectrum of LiH for the first 150 energy levels, indicating some agreement for the lowest-lying eigenstates but rapidly diverges. The inset plot shows the SSE spectrum (blue) and the true spectrum (orange), extending into higher energy levels.}
    \label{fig:LiH_spectrum}
\end{figure}

As the generalised eigenvalue problem (\cref{eq:generalised_eigenvalue_problem}) outputs an entire spectrum of energies within the subspace, it is interesting to compare the spectrum produced by SSE and the true spectrum of the Hamiltonian. In \cref{fig:LiH_spectrum} we show the energy difference between the two spectra for LiH for a run of SSE with $K=2000$ and without shot noise from the final state output by the variational algorithm in \cref{fig:LiH_initStates}. The ground state energy difference found is approximately $\Delta E=10^{-4}$, with the inset showing the two spectra individually (SSE in blue, true spectrum in orange). We can see that for the first $\sim 80$ eigenstates, the energies are in strong agreement with an energy difference of $\Delta E \lesssim 10^{-2}$. This falls off rapidly as we move to higher energy eigenstates. The agreement at the low energy level suggests the possibility of further extending our techniques to applications involving excited states.

\section{Conclusion}
In this work we present shadow subspace expansion (SSE) which leverages classical shadow data to perform high-dimensional subspace expansions for estimating ground state energies of quantum systems. We numerically investigate our approach by finding the ground state of a 14-qubit spin chain as well as  obtaining excited state energies of a LiH Hamiltonian as relevant in quantum chemistry.

We demonstrate that our method remains effective for subspace expansions of dimension up to the thousands -- which is beyond the size limitations of previous approaches by orders of magnitude. We observe more than one order of magnitude error reduction when using observable expectation values under a typical level of shot noise $\propto 10^{-3}$, while indeed further error reduction can be achieved by further suppressing shot noise (increasing the number of snapshots or circuit repetitions).

We specifically considered simple local Clifford shadows for estimating expansion operators as local Pauli strings and found that our method is robust against relatively small overlaps with the true ground state and, indeed, outperforms direct energy estimation (using the available expected values) in all examples. However, when the initial state is already very close to the true ground state, then we observe that an increased number of shadow snapshots are needed to obtain improvements beyond direct energy estimation. 
Furthermore, we demonstrate that the approach is also robust against gate noise in the state-preparation circuits.

Based on the potential advantages outlined above, when one performs ground-state energy estimation using classical shadows, our approach can be additionally performed fully in post processing at no quantum cost and can potentially further reduce errors in the energy estimates; Of course, in the worst case our approach yields no improvement in which case one still has access to direct energy estimate.

Given SSE is particularly effective when the initial state has relatively small overlap with the true ground state a potentially powerful application could be the following: if we are given classical shadows of the ground state of a given Hamiltonian, we can use this as an initial state for SSE to probe the ground state energy of slightly modified Hamiltonians, enabling us to probe the entire nearby energy landscape from the shadow data of one single state. Furthermore, in our numerical simulations we have primarily focused on estimating ground state energies, however, as we note in \cref{sec:procedure}, our method can directly be used for estimating other observables from the ground states.

Besides performing expansion of a given state within a subspace, one can also use post-processing techniques to project the state into a given subspace using techniques like symmetry verification~\cite{mcardleErrorMitigatedDigitalQuantum2019,bonet-monroigLowcostErrorMitigation2018}, in which the symmetry can arise from the target problems or qubit encoding~\cite{mccleanDecodingQuantumErrors2020}. Ref.~\cite{huLogicalShadowTomography2022} have applied classical shadows to symmetry verification in the context of quantum error correction codes and similar ideas are also discussed in Ref.~\cite{jnaneQuantumErrorMitigated2024}. It will be interesting to see if classical shadows can be applied to more general techniques that incorporate both symmetry verification and subspace expansion~\cite{caiQuantumErrorMitigation2021} as discussed in \cref{sec:symmetry}. 

Many further improvements are possible for SSE. The current version of SSE is still very susceptible to shot noise due to the finite number of shadow snapshots. This can be potentially alleviated by studying alternative shadow schemes and expansion bases, and also better ways to select the relevant expansion bases to perform the regularisation.

\begin{acknowledgments}
The authors thank Matthew Goh and Hamza Jnane for helpful comments.
Z.C. is supported by the Junior Research Fellowship from St John’s College, Oxford.
B.K. thanks the University of Oxford for
a Glasstone Research Fellowship and Lady Margaret Hall, Oxford for a Research Fellowship.
The numerical modelling involved in this study made
use of the Quantum Exact Simulation Toolkit (QuEST)~\cite{jonesQuESTHighPerformance2019}, and the recent development
QuESTlink~\cite{jonesQuESTlinkMathematicaEmbiggened2020} which permits the user to use Mathematica as the
integrated front end, and pyQuEST~\cite{meisterPyQuESTPythonInterface2024} which allows access to QuEST from Python.
We are grateful to those who have contributed
to all of these valuable tools. 
The authors also acknowledge supports from the
EPSRC projects Robust and Reliable Quantum Computing (RoaRQ, EP/W032635/1),
Software Enabling Early Quantum Advantage (SEEQA, EP/Y004655/1)
and the EPSRC QCS Hub grant (agreement No. EP/T001062/1).
\end{acknowledgments}

\appendix

\section{Symmetry Considerations} \label{sec:symmetry}

It may be desirable for the subspace expansion to find the lowest energy state with a certain symmetry, that is, within a subspace of the Hilbert space which has the correct eigenvalue of a symmetry operator $A$. For example, we may want to restrict our solution to the space with the correct number of electrons in a quantum chemistry application, to ensure the ground state energy estimate is not corrupted by non-physical states with the wrong number of electrons. \\
If the state rho $\rho$ has the correct value of this symmetry, it is straightforward to modify the subspace expansion to only choose operators that commute with the symmetry operator $\commutator{G_i}{A}=0$. \\
However, when our state $\rho$ does not itself lie within the symmetry subspace we must instead use post-processing to mitigate for the component with the wrong symmetry. One way of achieving this is to perform an analogue of symmetry verification \cite{bonet-monroigLowcostErrorMitigation2018}, and use existing techniques and perform an `internal' subspace expansion to individually mitigate all the matrix elements required \cite{caiQuantumErrorMitigation2021,jnaneQuantumErrorMitigated2024}. However, this requires increasing the weight of all observables measured by the weight of the symmetry operator(s), and may produce a large overhead due to the exponentially scaling shot requirements for higher weight observables.

There is an alternative method which we describe for the case of a single symmetry operator $A$ (although it is easily generalised). We can additionally determine the elements of the matrix

\begin{equation}
    \boldsymbol{A}_{ij} = \text{Tr}(G_i A G_j \rho)
\end{equation}

and before solving for the ground state energy, solve the generalised eigenvalue problem for the symmetry operator:

\begin{align} \label{eq:symmetry_generalised_eigenvalue_problem}
    \boldsymbol{A} \vec{w} = A_{\vec{w}}\boldsymbol{S}\vec{w}
\end{align}

In order to extract the subspace of the $\{G_j \ket{\psi} \}$ that (at least approximately) has the correct eigenvalue of the symmetry operator, and this can then be made the input for energy eigenvalue problem by first performing transformations analogous to \cref{eq:matrix_transform}, which will then produce a better estimate of the lowest energy within the correct symmetry subspace.

\section{Identifying good observables through time dependence} \label{sec:shadow_spectroscopy_observables}

We present here an alternative method for finding the $K$-significant observables based on Algorithmic Shadow Spectroscopy~\cite{chanAlgorithmicShadowSpectroscopy2023}. Given an initial state $| \psi \rangle = \sum_{k = 1}^d c_k \ket{ \psi_k }$
we time evolve it to obtain the quantum state $\ket{\psi(t)} := e^{- i t \mathcal{H}} \ket{\psi}$.
Measuring the expected value of any observable $O$ the signal $ S(t)$ is the following function of time as
\begin{equation}
	S(t)  := \langle \psi(t) |  O |\psi(t) \rangle = \sum_{k,l = 1}^d  I_{kl} \, e^{-i t (E_l - E_k) }.
\end{equation}
Here, the intensity carries all dependency on the observable as $I_{kl} = c_k^* c_l   \langle \psi_k | O | \psi_l \rangle$.

It follows that each signal has a phase-shifted cosinusoidal time dependence as 
\begin{equation}\label{eq:evol}
	S(t) = %\mathrm{const}  + \sum_{\substack{k,l=1\\k <l}}^q \mathrm{Re}[   I_{ikl}  e^{-i t (E_l - E_k) }]
	\sum_{k=1}^d   |c_k|^2  \langle \psi_k | O | \psi_k \rangle + \sum_{k<l}^d |I_{kl}| \cos[ t (E_l - E_k) + \phi_{kl}].
	%\nonumber
\end{equation}
Here we have used that $I_{kl}= |I_{kl}| e^{- i \phi_{kl}} = c_k^* c_l  \langle \psi_k | O | \psi_l \rangle$.

Assuming a low-energy initial state that has a dominant overlap with the ground state via $c_1 \gg c_k$ with $k>1$
we can approximate the signal by dropping all terms non-linear in $c_k$ as
\begin{equation}\label{eq:evol_2}
	S(t) \approx %\mathrm{const}  + \sum_{\substack{k,l=1\\k <l}}^q \mathrm{Re}[   I_{ikl}  e^{-i t (E_l - E_k) }]
	  |c_1|^2  \langle \psi_1 | O | \psi_1 \rangle + \sum_{k=2}^d |I_{1,k}| \cos[ t (E_1 - E_k) + \phi_{1,k}].
	%\nonumber
\end{equation}
In order to tell whether an observable $O$ can induce a transition between the ground
state and an excited state as $\langle \psi_k | O | \psi_1 \rangle$, it suffices to verify that the signal $S(t)$ is constant.
As we need not estimate the frequency of the signal $\cos[ t (E_1 - E_k) + \phi_{1,k}]$, it suffices to perform
a relatively short time evolution to obtain the signal $S(t)$ and apply a Ljung-Box test to verify whether 
the signal is statistically significantly different from shot noise. The approach is robust to algorithmic
errors in the time evolution since the frequency and the phase shift $\cos[ t (E_1 - E_k) + \phi_{1,k}]$
need not be estimated.

\section{Proof of \texorpdfstring{\cref{eq:error_prop}}{Noise Propagation}} \label{app:proof_propagation}
As detailed in the main text, we consider the solution to the generalised eigenvalue equation
as the eigenvalue equation for the matrix $M = \matinv	\boldsymbol{H}$
as
\begin{align} 
	\matinv	\boldsymbol{H} \vec{w} = E_{\vec{w}} \vec{w}.
\end{align}
We denote entries of the $q^{th}$ column vector of $M$ as $u^{(q)}$ which is equivalent to $u^{(q)} = \matinv v^{(q)}$,
where $v^{(q)}$ is the $q^{th}$ column vector of the Hamiltonian matrix $\boldsymbol{H}$.
Lemma~2 in Ref~\cite{vanstraatenMeasurementCostMetricAware2021} quantifies the
shot noise propagation in this inverse-matrix-vector
equation $u^{(q)} = A^{-1}v^{(q)} $ in terms of the total variance, i.e., the expected
Euclidean distance squared, of the vector $u^{(q)}$ due to shot noise 
as  $\epsilon_q = \sum_k \var{u^{(q)}_k}$.

The following bound was obtained for the shot-noise propagation
\begin{align*}
	&\epsilon_q^{2} = \sum_{k, l = 1}^{\nu} a_{k l} \myvar \big \{[\mat]_{k l}\big\}  + \sum_{k = 1}^{\nu} b_{k} \myvar [v^{(q)}_k], \\
	& \textnormal{where} \ \ a_{k l} := \sum_{i,j = 1}^{\nu} [\matinv]_{ik}^{2} [\matinv]_{l j}^{2} [v^{(q)}_{l}]^{2},\\
	& \textnormal{and} \ \ b_{k} := \sum_{l = 1}^{\nu} \big \{[\matinv]_{k l} \big \}^{2}.
\end{align*}
A bound on  the following coefficients was proved as
\begin{equation}
	a_{k l} \leq \lVert v^{(q)} \rVert_\infty^{2} \sum_{i,j = 1}^{\nu} [\matinv]_{ik}^{2} [\matinv]_{l j}^{2} 
\end{equation}

We assume that the shot noise in the matrix entries $[\mat]_{k l}$ and $[\boldsymbol{H}]_{kl}$ is independent (zero covariance).
We expect the bounds on the shot-noise propagation to be loose, and we are primarily interested
in establishing the scaling of the error rather than a guaranteed bound.

We can apply the variance bound for classical shadows given that in the present work we assume
the matrix entries $\myvar \big \{[\boldsymbol{S}]_{k l}\big\}$ are estimated from classical shadows.
Given all matrix entries are of the form of \cref{eqn:matrix_elem}, the variance is upper bounded as
\begin{equation}
	\myvar \big \{[\boldsymbol{S}]_{k l}\big\} \leq N_s^{-1} \, 3^{2w'}.
\end{equation}
Above we used the largest weight $w'$ among the expansion operators and $N_s$ is the number of shots used.

Similarly, we can bound the variance of the Hamiltonian matrix entries as
\begin{equation*}
	\myvar  [v^{(q)}_k] =   \myvar \big \{ [ \boldsymbol{H} ]_{k  q} \big\} \leq N_s^{-1} \, 3^{w + 2w'},
\end{equation*}
where we used the largest weight among the Pauli terms in the Hamiltonian.

Substituting this back into the error propagation formula we obtain
\begin{equation}
\epsilon_q^{2} \leq  N_s^{-1} \, 3^{2w'} \sum_{k, l = 1}^{\nu} a_{k l}  + N_s^{-1} \, 3^{w + 2w'} \sum_{k = 1}^{\nu} b_{k}.
\end{equation}
Substituting back the explicit form of $b_{k}$ and our upper bound for $a_{k l}$ we finally obtain
\begin{equation}
	\epsilon_q^{2} \leq  N_s^{-1} \, 3^{2w'} \lVert v^{(q)} \rVert_\infty^{2} \lVert \matinv \rVert_2^4  + N_s^{-1} \, 3^{w + 2w'} \lVert \matinv \rVert_2^2,
\end{equation}
where $\lVert \cdot \rVert_2$ is the Hilbert-Schmidt or Frobenius norm of a matrix.

We have so far quantified the total variance in the $q^{th}$ column vector of $M$. We define the total variance
in $M$ as $\epsilon_{M}^2 = \sum_{q=1}^{K} \epsilon_q^{2} = \sum_{q k} \myvar \big \{ [M]_{kq} \big \} $
which is the expected Hilbert-Schmidt distance in the matrix $M$ due to shot noise.
Using that 
\begin{equation}
\sum_{q=1}^{K} \lVert v^{(q)} \rVert_\infty^{2}  
\leq K \max_{kl}  [\boldsymbol{H}]_{kl} \leq   K  \lVert H_{\mathrm{tot}} \rVert_\infty,
\end{equation}
where $\lVert H_{\mathrm{tot}} \rVert_\infty$ is the absolute largest eigenvalue of the problem Hamiltonian.
We finally obtain the bound 
\begin{equation}
	\epsilon_{M}^2 \leq  N_s^{-1} \, 3^{2w'}  K  \Big[   \lVert H_{\mathrm{tot}} \rVert_\infty  \lVert \matinv \rVert_2^4  +3^{w } \lVert \matinv \rVert_2^2  \Big ].
\end{equation}

\begin{figure}
	\includegraphics*[width=\linewidth]{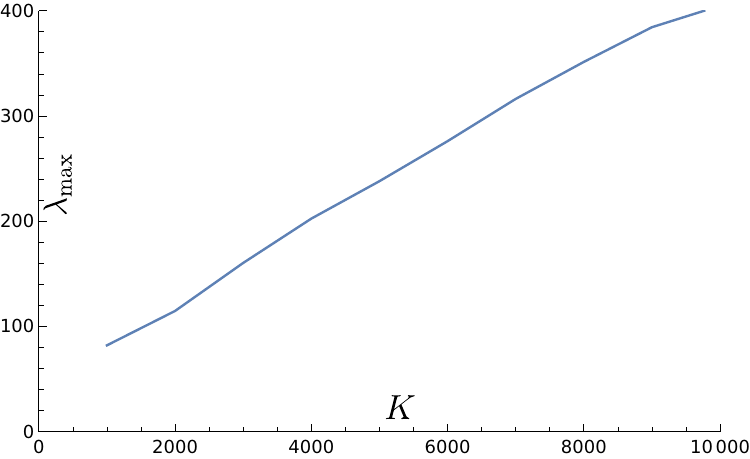}
	\caption{Demonstration of the largest eigenvalue of $\boldsymbol{S}$ increasing linearly with $K$, as is expected for a matrix of overlaps $\boldsymbol{S}_{ij} = \braket{\psi_i}{\psi_j}$, where the states are mostly similar to each other (consisting of only small local changes). For example in the extreme case where we take all $G_i=I$, then the largest eigenvalue is $K$.}
	\label{fig:largestSV}
\end{figure}

\section{Variance of Local Clifford Shadow} \label{sec:variance_local_clifford_shadow}
Suppose we are given the ability to prepare a given state $\rho$, and we want to construct its classical shadow using local Clifford twirling. In a given circuit run, we will prepare the state $\rho$ and then perform measurements corresponding to a set of random local Pauli operators $\hat{\vec{G}}$, which give the set of random outcomes $\hat{\vec{g}} \in \{\pm 1\}^{\otimes N}$. The output state $\pket*{\hat{\vec{G}},\hat{\vec{g}}}$ is stabilised by the set of Pauli generators $\hat{\vec{g}} \odot \hat{\vec{G}}$ where $\odot$ is element-wise multiplication, and we will call this a \emph{projection} of the input state $\rho$. If we use $\mathcal{D}$ to denote the channel where each qubit is locally depolarised with $2/3$ probability, then by applying the inverse of this channel on the output state, $\pket{\hat{\rho}} = \mathcal{D}^{-1}\pket*{\hat{\vec{G}},\hat{\vec{g}}}$, will give a classical snapshot of the input state $\rho$ and the collection of them form the classical shadows of $\rho$. We can obtain the expectation value of a Pauli observable $O$ with respect to the input state $\rho$ by taking the expectation value of $O$ with respect to these snapshots:
\begin{align*}
    \pbraket{O}{\rho}  = \expect{\pbraket{O}{\hat{\rho}}} = \expect{\pbra{O}\mathcal{D}^{-1}\pket*{\hat{\vec{G}},\hat{\vec{g}}}}
\end{align*}

Using the fact that $\pbra{O}\mathcal{D}^{-1}\pket*{\hat{\vec{G}},\hat{\vec{g}}} = 3^{w}\pbraket*{O}{\hat{\vec{G}},\hat{\vec{g}}}$ where $w$ is the weight of $O$, we can estimate $\pbraket{O}{\rho}$ via two steps.
Step one is to obtain the random variable $\hat{Q} = \pbraket*{O}{\hat{\vec{G}},\hat{\vec{g}}}$ through post-processing for each snapshot. Two possible scenarios can arise:
\begin{enumerate}[leftmargin=*]
\item $O \in \expval*{\hat{\vec{G}}}$, i.e. $O$ can be obtained through multiplying the set of local Pauli measurements in the given snapshot, or in another word, $+O$ or $-O$ is a stabiliser of the snapshot. This happens with the probability $3^{-w}$ since each local Pauli measurement is chosen randomly from $\{X, Y, Z\}$. In this case, the output random variable $\hat{O}$ follows the same distribution as directly measuring $O$ on state $\rho$. Thus, we have $\expect{\hat{O}} = \pbraket{O}{\rho}$. 
\item $O \not\in \expval*{\hat{\vec{G}}}$ which happens with the probability $1 - 3^{-w}$. In this case, we simply have $\pbraket*{O}{\hat{\vec{G}},\hat{\vec{g}}} = 0$. 
\end{enumerate}
From the description above we have: 
\begin{align*}
    \expect{\hat{Q}} &= 3^{-w}\expect{\hat{O}} + (1-3^{-w}) 0 = 3^{-w}\expect{\hat{O}}\\
\expect{\hat{Q}^2} &= 3^{-w}\expect{\hat{O}^2} + (1-3^{-w}) 0^2 = 3^{-w}\\
\var{\hat{Q}} &=\expect{\hat{Q}^2} - \expect{\hat{Q}}^2 =  3^{-w}  - 3^{-2w}\expect{\hat{O}}^2
\end{align*}
    
Step two is rescaling the result obtained from $\hat{Q}$ by $3^w$, which gives rise to the estimator $\hat{R} = 3^w\hat{Q}$. This gives:
\begin{align*}
    \expect{\hat{R}} = 3^{w}\expect{\hat{R}} = \expect{\hat{O}} = \pbraket{O}{\rho}
\end{align*}
as expected, so the shadow estimator $\hat{R}$ is an unbiased estimator of $\pbraket{O}{\rho}$. Its variance is given as:
\begin{align*}
\var{\hat{R}} &= 3^{2w}\var{\hat{Q}} = 3^{w}  - \expect{\hat{O}}^2.
 \end{align*}

\section{Prior Work on Shot Noise Propagation\label{app:shot_noise2}}
As derived in the Appendix of \cite{yoshiokaGeneralizedQuantumSubspace2022}, when solving
\begin{align*} 
    \boldsymbol{H} \vec{w} = E_{\vec{w}}\boldsymbol{S}\vec{w},
\end{align*}
the shot noise in $\boldsymbol{H}$ and $\boldsymbol{S}$, denoted as $\delta \boldsymbol{H}$ and $\delta \boldsymbol{S}$, will also lead to noise in the calculated weight vectors and energies. On can show that
\begin{align*}
    \delta E_{\vec{w}} = \vec{w}^{\dagger} \left(\delta \boldsymbol{H} - E_{\vec{w}} \delta \boldsymbol{S}\right) \vec{w}
\end{align*}
As shown in \cite{yoshiokaGeneralizedQuantumSubspace2022}, this error is bounded by
\begin{align*}
    \delta E_{\vec{w}} \leq  \norm{\boldsymbol{S}^{-1}}_{\infty}  \left(\norm{\delta \boldsymbol{H}}_{\infty} + \abs{E_{\vec{w}}}\norm{\delta \boldsymbol{S}}_{\infty}\right).
\end{align*}
Furthermore, we can bound the energy as $\abs{E_{\vec{w}}} \leq \lVert H_{\mathrm{tot}} \rVert_\infty$ which yields the bound
\begin{align*}
	\delta E_{\vec{w}} \leq  \norm{\boldsymbol{S}^{-1}}_{\infty}  \left(\norm{\delta \boldsymbol{H}}_{\infty} +\lVert H_{\mathrm{tot}} \rVert_\infty  \norm{\delta \boldsymbol{S}}_{\infty}\right).
\end{align*}

The dependence on $\norm{\boldsymbol{S}^{-1}}_{\infty}$ means that truncation of the expansion basis is essential such that $\boldsymbol{S}$ is not close to being singular.

\section{Dependence of eigenvalues of \texorpdfstring{$\boldsymbol{S}$}{S} on \texorpdfstring{$K$}{K}} \label{sec:largest_SV_with_K}

See \cref{fig:largestSV} for the dependence of the largest eigenvalue of $\boldsymbol{S}$ with $K$.

\bibliography{ref}
\end{document}